\documentclass[a4paper]{iopart}

\usepackage{graphicx}

\begin{document}

\title{Theoretical model for ultracold molecule formation 
       via adaptive feedback control}

\date{\today}

\author{Ulrich Poschinger, Wenzel Salzmann, Roland Wester, and
  Matthias Weidem\"uller
}
\address{Physikalisches Institut der Universit\"at Freiburg,
  Hermann-Herder-Str. 3, 79104 Freiburg, Germany
}
\eads{\mailto{roland.wester@physik.uni-freiburg.de}}
\eads{\mailto{m.weidemueller@physik.uni-freiburg.de}}

\author{Christiane P. Koch and Ronnie Kosloff
}
\address{Department of Physical Chemistry and
  The Fritz Haber Research Center,
  The Hebrew University, Jerusalem 91904, Israel
}

\eads{\mailto{ckoch@physik.fu-berlin.de}}


\begin{abstract}
  We theoretically investigate pump-dump photoassociation of ultracold
  molecules with amplitude- and phase-modulated femtosecond laser pulses. For
  this purpose a perturbative model for the light-matter interaction is
  developed and combined with a genetic algorithm for adaptive feedback
  control of the laser pulse shapes. The model is applied to the formation of
  $^{85}$Rb$_2$ molecules in a magneto-optical trap. We find that optimized
  pulse shapes may maximize the formation of ground state molecules in a
  specific vibrational state at a pump-dump delay time for which unshaped
  pulses lead to a minimum of the formation rate. Compared to the maximum
  formation rate obtained for unshaped pulses at the optimum pump-dump delay,
  the optimized pulses lead to a significant improvement of about 40\% for the
  target level population. Since our model yields the spectral amplitudes and
  phases of the optimized pulses, the results are directly applicable in pulse
  shaping experiments.
\end{abstract}

\maketitle

\section{Introduction}
\label{sec:intro}

One of the challenges in contemporary atomic and molecular physics is the
making of ultracold molecular samples \cite{ElectronDipoleMoment, BCSPairing,
QCwithmolecules}. Direct cooling techniques for molecules have not yet reached
the ultracold regime ($T<1\,$mK). Instead, ultracold atoms are assembled to
molecules using magnetic Feshbach resonances \cite{GrimmFeshbach} or
photoassociation \cite{FirstPAProposal,FirstPA}. In the latter case, typically
a continuous-wave (cw) laser excites a pair of colliding atoms into a
long-range molecular state. Subsequent spontaneous emission leads to molecules
in their electronic ground state. In view of prospective applications, stable
ultracold molecules, i.\ e.\ molecules in their absolute rotational,
vibrational and electronic ground state are required. However, molecules
formed with Feshbach resonances are created in the highest bound vibrational
level \cite{GrimmFeshbach}. Photoassociation with a cw laser followed by
spontaneous emission forms molecules in a range of vibrational levels
\cite{FirstPAGroundstate}. In this case, the radiative lifetime of the excited
electronic state ultimately represents the rate-determining step in ground
state molecule formation. When applied to a Bose-Einstein condensate (BEC),
the incoherent nature of the spontaneous decay furthermore destroys the
coherence of the BEC and hence the condensate itself \cite{McKenzie}. This may
be avoided by stimulated Raman transitions \cite{vanhaecke2004:epd}. Recently,
also light-driven Rabi oscillations between atoms and molecules have been
observed \cite{ryu2006:arxiv}. Using cw photoassociation in combination with
nanosecond pulsed Raman transitions, the formation of RbCs molecules in the
$v=0$ level of the ground electronic state has been achieved
\cite{sage2005:prl}.

\begin{figure}[tb]
\centering
\includegraphics[width=0.7\textwidth]{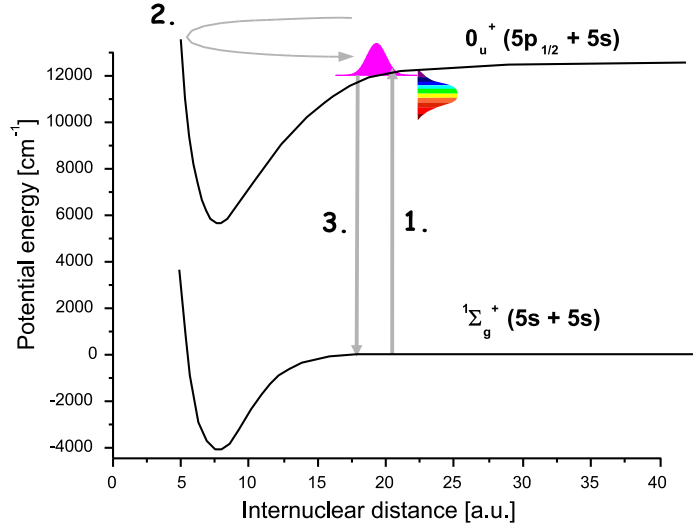}
\caption{The pump-dump scheme for the production of
  ultracold molecules with short laser pulses. A wavepacket is created
  in the excited state by a pump pulse (1) and subsequently propagates
  (2). A dump pulse transfers population to the target bound ground
  state level after a suitable time delay (3).
}
\label{PumpDumpIntroduction}
\end{figure}

An alternative approach to ultracold molecule formation is the
photoassociation with time-dependent laser fields of femtosecond to picosecond
pulse durations. The conceptually simplest pulsed photoassociation scheme is
based on a two-pulse pump-dump population transfer from the atomic continuum
to a bound molecular level \cite{WavepacketMachholm}, as shown in
\fref{PumpDumpIntroduction}. The pump pulse excites a molecular wavepacket of
the vibrational states within its bandwidth, which then propagates until it is
dumped to the ground state by a second laser pulse. By modifying the delay
time between the two pulses, the transition to one or several desired target
vibrational levels in the ground state can be enhanced by quantum
interference. Here, the production rate is not limited by the radiative
lifetime time of the intermediate state, but by the much shorter vibrational
period, by the Franck-Condon factors and ultimately by the unitarity limit of
the scattering rate \cite{McKenzie,WesterSaturation}. The particular advantage
of this approach is that the pump-dump delay time and the pulse shapes yield
additional control knobs to prepare specific product states, which is
extensively explored in the field of adaptive feedback control with shaped
femtosecond pulses \cite{TeachingLaserToControlMolecules,BrixnerReview}. A
number of theoretical studies have investigated the use of chirped laser
pulses for excitation \cite{ChirpedPAValaKosloff,ChirpedPA1,ChirpedPA2} and
stabilization \cite{PumpDumpKoch,ChristianeD1} of ultracold molecules. Even
larger control over the molecule formation process might be obtained by shaped
laser pulses. Laser pulses which optimize the formation of ultracold molecules
from an atomic BEC have been calculated using optimal control theory
\cite{hornung2002:pra}. In many cases the complexity of the pulse shapes
resulting from optimal control calculations requires a careful transformation
into experimentally achievable pulse shapes \cite{HornungPRA02a, MancalCPL02}.

In this paper, adaptive feedback control of photoassociation with femtosecond
laser pulses is investigated theoretically. For this purpose the properties of
typical femtosecond oscillator pulses and a realistic pulse shaping device are
combined with a closed loop genetic algorithm for pulse shape
optimization. The resulting pulse shapes are obtained as amplitudes and phases
in the frequency domain, similar to the case in closed-loop adaptive feedback
control experiments. The parameters of the calculations are chosen to
correspond to a situation, which is typical for state-of-the-art experiments
\cite{WenzelSeinPaper}. A two-channel model is employed, which simplifies the
complex coupled-channel dynamics, but retains the important features of
pump-dump photoassociation. The goal of this study is to qualitatively
determine to which extent the formation process of ground state molecules can
be controlled by realistic shaped laser pulses.

A first experiment on photoassociation with femtosecond laser pulses has been
carried out on hot mercury atoms \cite{marvet1995:cpl}. In the ultracold
regime, electronically excited Na$_2$ molecules have been formed using
picosecond pulses near-resonant to the sodium D1 transition, and probed with a
subsequent laser pulse \cite{fatemi2001:pra}. Chirped nanosecond laser fields
have been employed to enhance inelastic collision rates in an ultracold gas
\cite{wright2005:prl}. Also the effect of the chirp of femtosecond pulses on
the dissociation of ultracold Rb$_2$ molecules has been investigated
\cite{brown2006:prl}. Recently, adaptive feedback control with shaped
femtosecond pulses has been applied in our group to optimize the dissociation
of ultracold Rb$_2$ molecules from a magneto-optical trap
\cite{WenzelSeinPaper}. We envision that the work presented in this paper will
be helpful for the first experimental demonstration of ultracold ground state
molecule formation using pulsed photoassociation, which has not been achieved
to date.

The paper is organized as follows: the model of two colliding atoms
interacting with the laser fields is described in \sref{subsec:system}. A
perturbative treatment of the interaction with the field is employed in
\sref{subsec:shapedpulses} to obtain a model for the pulse shaping device. Its
optimization with a genetic algorithm is outlined in \sref{subsec:GA}. The
results are presented in \sref{sec:results}: The validity of the perturbative
model is tested in \sref{subsec:valid} by comparison to full, non-perturbative
quantum calculations. In \sref{subsec:optdet} and \sref{subsec:control} the
results for optimal laser pulses for pulsed photoassociation, as obtained
from adaptive feedback control are presented and analyzed.

\section{Model for the photoassociation process}

\subsection{Interaction of two colliding atoms with a laser field}
\label{subsec:system}
A colliding atom pair subject to a laser pulse is considered. Two electronic
states coupled by the radiation field are taken into account with the
assumption of the rotating wave approximation. The center-of-mass motion of
the two atoms is assumed to be decoupled from the internuclear dynamics.  The
time-dependent Schr\"odinger equation then reads
\begin{equation}
i\hbar\;\frac{\partial}{\partial t}\left( \begin{array}{c} \psi_g(R,t) \\
\psi_e(R,t)\end{array} \right)=\hat{H} \left( \begin{array}{c} \psi_g(R,t) \\
\psi_e(R,t)\end{array} \right)
\label{eq:CentralTDSE}
\end{equation}
with
\begin{equation}
\hat{H}=\left(
\begin{array}{cc} \hat{T}+V_g(R) & \hbar \Omega(t) \\
\hbar \Omega^{*}(t) & \hat{T}+V_e(R)-\Delta \end{array} \right) \,.
\label{eq:CentralHamil}
\end{equation}
The wavefunction consists of one component for the ground state,
$\psi_g(R,t)$, and one component for the excited state, $\psi_e(R,t)$. The
diagonal elements of the Hamiltonian are given by the kinetic energy operator
$\hat{T}$ and the potential energy curves for the respective states
$V_{g/e}(R)$ which depend on the internuclear separation $R$. The zero of
energy is chosen to correspond to the ground state dissociation limit. Within
the rotating wave approximation, $V_e$ is shifted by the detuning $\Delta$
with respect to $V_g$ where $\Delta=\hbar (\omega_{ge} -
\omega_L)$. $\omega_{ge}$ denotes the atomic transition frequency and
$\omega_L$ the carrier frequency of the laser. The off-diagonal elements of
the Hamiltonian are given by the complex time-dependent Rabi frequency
$\Omega(t)$, which is related to the electric field $\tilde \mathcal{E}(t)$ by
$\Omega(t)=\mu \tilde \mathcal{E}(t)/\hbar$. Here, $\mu$ denotes the dipole
matrix element of the two molecular electronic states under consideration. In
the following, the $R$-dependence of this matrix element is neglected and its
asymptotic value is employed, $\mu=2.537\cdot10^{-9}\,$Cm for the Rb D1 line.
The laser pulse is assumed to be derived from a Gaussian transform-limited
pulse sent through a pulse shaping device as detailed in \sref{subsec:GA}.

The two channels under consideration for the calculations are the singlet
ground state $X^1\Sigma_g(5s+5s)$ and the excited state $0_u^+(5s+5p_{1/2})$
of $^{85}$Rb$_2$. The wavelength of the asymptotic (atomic) transition
corresponds to 794.9\,nm. The potential energy curves are derived from
\textit{ab initio} data \cite{RbPotentialsPark} at short distances and
long-range dispersion potentials $(-C_3/R^3)-C_6/R^6-C_8/R^8$ for distances
larger than the LeRoy-Bernstein radius. The coefficients for the $5s+5s$
asymptote are found in \cite{MartePRL02}, while the coefficients for the
$5s+5p$ asymptote are taken from \cite{GuterresPRA02}. The Hund's case (c)
representation of the potential energy curves is obtained by diagonalizing the
Hund's case (a) potentials coupled by spin-orbit interaction
\cite{SpiegelmannPot}. The spin-orbit coupling matrix element is set to the
atomic value. The $R$-dependence and the resonant character of the spin-orbit
coupling are neglected. The latter leads to strong perturbations in the
vibrational spectrum \cite{AmiotPRL99} providing an efficient stabilization
mechanism for pump-dump photoassociation \cite{ChristianeD1}. However, its
treatment would require a three-channel model, and the focus of this study is
on the qualitative understanding of possible control over the wavepacket
dynamics.

The Hamiltonian \eref{eq:CentralHamil} is represented on a grid using the
mapped Fourier grid method~\cite{MappedFGHKokoouline,MappedFGHWillner}. A grid
of 1024 points extending up to 27000 Bohr radii is chosen. This size is
necessary to obtain a reliable representation of continuum scattering states
\cite{ChirpedPA2}. The binding energies $E^{g/e}_v$ and wavefunctions
$\varphi^{g/e}_v(R)$ of the vibrational levels are computed by diagonalizing
the Hamiltonian \eref{eq:CentralHamil} with $\Omega(t)=0$. Franck-Condon
factors are then obtained as scalar products, $|\langle v | w \rangle|^2 =
\int dR |{\varphi^e_{v}(R)}^*\varphi^g_w(R)|^2$. To test the validity of the
perturbative model which is explained in the next section, the time-dependent
Schr\"odinger equation \eref{eq:CentralTDSE} is solved with the Chebyshev
propagator method \cite{KosloffTimeDependentMethods}.

The initial state of the colliding atom pair is chosen such as to describe an
ultracold atomic gas in a magneto-optical trap.  Due to the low temperatures
of about 100$\,\mu$K, only a narrow band of continuum states above the ground
state potential asymptote is occupied.  The considerations in this study focus
on a single initial continuum state with collisional angular momentum $l=0$,
i.e. an s-wave scattering state. This assumption is justified since the
continuum wavefunctions corresponding to temperatures $T \le 500\,\mu$K
possess an identical nodal structure at short and intermediate internuclear
distances. A thermal average over all possible initial states is only required
for absolute molecule formation rates, which is beyond the scope of the
present study.

\subsection{Model of the photoassociation process with shaped laser pulses}
\label{subsec:shapedpulses}
In the following a perturbative model of pump-dump photoassociation with
arbitrarily phase and amplitude modulated laser pulses is presented. Such
shaped laser pulses are prepared experimentally by passing the spectrally
dispersed pulses through a liquid crystal modulator array. In each pixel of
such a pulse shaper the phase and the amplitude of the corresponding frequency
component of the pulse can be computer controlled.

The photoassociation pulse creates a time-dependent wavepacket in the excited
state. After the pulse is off, the excited state wavefunction $\psi_e(R;t)$ is
given by a superposition of vibrational levels in the excited state. If the
interaction can be treated perturbatively, i.e. the Rabi angle, $\theta= \int
\Omega(t)\;dt$, is small compared to $\pi$, the pulse excites the
superposition of vibrational levels given by \cite{WavePacketWalmsley}
\begin{equation}
  \psi_e(R;t) 
  \propto 
  \sum_{v} a_{v} \;
  \mathcal{E}(\omega_v)
  \langle v | E \rangle \;
  \varphi_{v}(R) \;
  e^{-i E^e_{v} t/\hbar} \,,
  \label{pumpsuperp}
\end{equation}
where $\langle v | E \rangle$ is the overlap integral of the initial continuum
state $|E\rangle$ and the excited state vibrational level $v$; its square
corresponds to the free-bound Franck-Condon factors. The frequency spectrum of
the electric field of the laser pulse, $\mathcal{E}(\omega)$, is evaluated at
the transition frequency $\omega_v$ to the vibrational level $v$. The $a_v$
coefficients are introduced to represent the complex modulation coefficients
of a pulse shaper. This description assumes that the frequency components for
each contributing $v$ are individually resolved by the pulse shaper. As one
can experimentally only diminish spectral components, one obtains the
constraint $0\leq |a_v|\leq 1$. 

\Eref{pumpsuperp} is only valid, if the pulse does not contain significant
amplitude at frequencies close to resonance with the atomic transition. This
is due to the strongly increasing transition strength close to the atomic
resonance, which leads to a break-down of the perturbative treatment for
typical laser intensities. We estimate that the perturbative model is valid if
the detuning of the laser pulse spectrum with respect to the atomic transition
is large enough so that minimal excitation of the atomic transition
occurs. After a delay time $\Delta t$, a second or dump pulse
$\mathcal{E}'(\omega)$ is applied. This pulse transfers population from the
excited state to the ground state; it is assumed to be not shaped by a pulse
shaper. The resulting superposition of \textit{bound} ground state levels
reads
\begin{eqnarray}
\psi_g^\mathrm{bound}(R;t) &
\propto &
\sum_{v,w} a_{v} 
\mathcal{E}(\omega_v)
\mathcal{E}'(\omega_{v w}) \nonumber \\
& &
\quad\quad 
\langle E | v \rangle \langle v | w \rangle \; \varphi_w(R) 
e^{-\frac{i}{\hbar} E^g_w (t-\Delta t)}
e^{-\frac{i}{\hbar} E^e_{v} \Delta t}
\,.
\label{pumpsuperp2}
\end{eqnarray}
Note that this description is only valid for non-overlapping pulses, because
transient dynamics that occur during the excitation process are not included.
Therefore only delay times $\Delta t$ that are large compared to the duration
of the pump and dump pulses are chosen. In \eref{pumpsuperp2}, $v$ runs over
all bound vibrational levels of the excited electronic state and $w$ over all
bound levels of the ground state. The $\langle v | w \rangle$ are overlap
integrals between ground and excited state bound levels. The time delay
between the two pulses enters the expression through the phase factors $e^{-i
E^e_v \Delta t/\hbar}$ which corresponds to the propagation of the wavepacket
in the excited state. The population of an individual bound ground state level
$w$ is now given by
\begin{eqnarray}
| b_{w} |^2 &
= &
\sum_{v,v'} a_{v} a_{v'}^{*}
\mathcal{E}(\omega_v)
\mathcal{E}'(\omega_{v w})
\mathcal{E}^{*}(\omega_{v'})
\mathcal{E}'^{*}(\omega_{v' w}) \nonumber \\
 & &
\quad\quad 
\cdot 
\langle E | v \rangle \langle v | w \rangle \langle w | v'
\rangle \langle v' | E \rangle 
e^{-\frac{i}{\hbar} (E^e_{v}-E^e_{v'}) \Delta t}
\,.
\label{pumpsuperp3}
\end{eqnarray}
This double sum over the excited state vibrational levels $v$ and $v'$ can be
concisely written as a complex quadratic form in terms of the pulse modulation
coefficients $a_v$,
\begin{equation}
| b_{w} |^2 
=
\sum_{v,v'} 
a_{v'}^{*} \; M^{vv'}_{w}(\Delta t)\; a_{v}.
\label{PAtargetrate}
\end{equation}
The Franck-Condon matrix $M^{vv'}_{w}(\Delta t)$ contains the properties of
the system (overlap integrals and binding energies) and of the frequency
spectrum of the two laser pulses and their pulse delay.

It can be inferred from \eref{PAtargetrate} that the choice of modulating the
pump pulse and not the dump pulse is in fact arbitrary, since the same
structure of \eref{PAtargetrate} is obtained if the dump pulse or both pulses
are shaped. \Eref{PAtargetrate} serves two purposes: First, for a given target
level $w$, it can be used to obtain the optimal detuning of the pulses by
evaluating $|b_w|^2$ for different delay times $\Delta t$. Second, it allows
for the implementation of an optimization procedure as explained in the
following section.

\subsection{Genetic algorithm for the pulse optimization}
\label{subsec:GA}

In the following, the photoassociation pulse shall be optimized while the dump
pulse is assumed to be transform-limited. An optimization algorithm is
obtained by combining the analysis of the target state \eref{PAtargetrate}
with a spectral decomposition of the pulse. The photoassociation pulse which
yields the desired excited state wavepacket can in principle be spectrally
composed from a set of weighted delta functions at the transition frequencies
corresponding to the vibrational states $v$ \cite{WavePacketWalmsley}. This
spectrum of a few sharp structures leads to very long pulses in time domain,
which is unfavorable for our perturbative model of a pump dump scheme as
explained above. The photoassociation pulse is instead constructed from a
coherent superposition of a larger number of frequency components,
\begin{equation}
\hbar \Omega(t)
=
\mu 
\sum_{n=1}^{N_p} 
c_n
\mathcal{E}(\omega_n)
e^{i (\omega_n-\omega_L) t} \,,
\label{PulseTransform}
\end{equation}
which leads to a shorter pulse in the time domain. The frequency spectrum of
the pump and dump laser pulses before modifications by the pulse shaper is
assumed to be a Fourier limited Gaussian pulse $\mathcal{E}(\omega_n) \propto
\exp{[-(\omega_n-\omega_L)^2 / {2\Gamma^2}]}$ with a root-mean-square width
$\Gamma$. Each $\omega_n$ corresponds to a pixel in the pulse shaper mask,
$\omega_L$ is again the carrier frequency of the laser pulse. The pixel number
$N_p$ is set to 640 and the frequency spacing from pixel to pixel is set to
1.5\,cm$^{-1}$, which are numbers used in experimentally available pulse
shaping devices. This leads to a pulse shaper resolution that is smaller than
the vibrational level spacing for almost all the excited vibrational
levels. The pulse shaper coefficients $c_n$ are set equal to the desired
superposition coefficients $a_v$ from \eref{PAtargetrate}, where $v$ is the
excited state level for which $\hbar E_v$ is closest to $\omega_n$. The range
of vibrational levels which contribute to the population in the excited state
is identified by the non-vanishing matrix elements of $M^{vv'}_{w}$. In the
present case, this range consists of 150 levels ranging from $v=170$
($T_\mathrm{vib}=0.77\,$ps) to $v=320$ ($T_\mathrm{vib}=80.0\,$ps). The
discreteness of the spectrum \eref{PulseTransform} still leads to unwanted
replica effects in the time domain. The pulse is therefore multiplied with a
box function of the form $\exp{[-(t/T_{\rm cutoff})^m]}$ with $m=8$ and
$T_{\rm cutoff} = 3.33$\,ps; $t$ is the time relative to the maximum of the
pulse envelope. This provides a smooth switching on and off of the pulse.

The optimization is carried out with respect to the $a_v$ in
\eref{PAtargetrate}, corresponding to the pixels of a pulse shaping
device. The pump-dump delay time $\Delta t$ represents a fixed parameter for
the optimization process. It turns out that for the amplitude optimization it
is sufficient to set $a_v$ to either 0 or 1. For the phase $256$ possible
values are considered, leading to an overall 9-bit optimization of the
$a_v$. The quality of the pulses is determined by the fitness factor $Q$ which
is chosen as
\begin{equation}
Q=\frac{|b_w|^2}{\sqrt{\langle t^2 \rangle}} \,.
\end{equation}
Here, $\langle t^2 \rangle$ is a measure of the pulse duration,
\begin{equation}
\langle t^2 \rangle=\frac{\int t^2\;\Omega(t)\;dt}{\int
\Omega(t)\;dt} \,.
\end{equation}
It is included in the definition of the fitness factor in order to keep the
pulses as short as possible. This is not a physical constraint but a
requirement of the perturbative model, because the duration of the
pulses is immensely increased by the modulation: Already a linear frequency
chirp stretches the pulse in time. The employed perturbative model, however,
requires non-overlapping pump and dump pulses, as explained above.

The optimization of the formation of molecules in a specific ground state
level $w$ is carried out by means of an evolutionary algorithm. This
corresponds directly to a close-loop experiment. For the calculations a very
basic evolutionary algorithm, \textit{SimpleGA} from the \textit{GALib}
package \cite{ga_mit:html}, is employed. Populations of 20 individuals are
used, and the number of generations is set to 1000. This is sufficient to
obtain convergent results. The single-bit mutation and single-bit crossover
probabilities are chosen as 0.005 and 0.8, respectively.

\section{Results}
\label{sec:results}

\subsection{Validity of the pulse shaper model}
\label{subsec:valid}

The pulse shaper model of \eref{PAtargetrate} is derived for the case of a
small perturbation of the atom pair by the laser field. Higher order
processes, in particular Rabi cycling during the pump or dump pulses, are
neglected. By carrying out calculations for a single unshaped pump pulse with
varying the peak intensity of the pulse, the regime of validity of the
perturbative approximation is analyzed.

\begin{figure}[tbp]
\center
\includegraphics[width=0.7\textwidth]{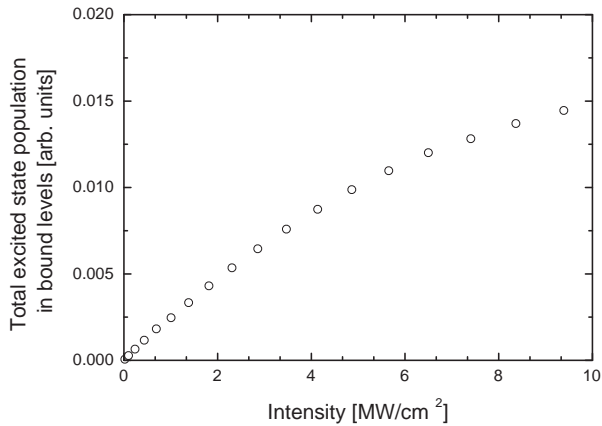}
\caption{Dependency of the excitation probability on the peak intensity of the
  pump pulse for a fixed pulse width of 100\,fs FWHM. The linear dependence on
  intensity below 2\,MW/cm$^2$ shows the validity of a perturbative
  treatment.}
\label{IntScan}
\end{figure}

In \fref{IntScan} the probability to excite an atom pair to the excited state
is shown as a function of the pulse's peak intensity. These results are
obtained by solving the time-dependent Schr{\"o}dinger equation where the
field is treated non-perturbatively. \Fref{IntScan} clearly shows a linear
dependence of the excitation probability for small intensities
$<4\cdot10^6$\,W/cm$^2$. For these low intensities the effect of the laser
field can be treated in first order perturbation theory. For larger pulse
energies the excitation starts to saturate, which indicates the transition to
a non-perturbative regime. This is attributed to Rabi cycling and strong
off-resonant excitations.

\begin{figure}[tbp]
\center \includegraphics[width=0.6\textwidth]{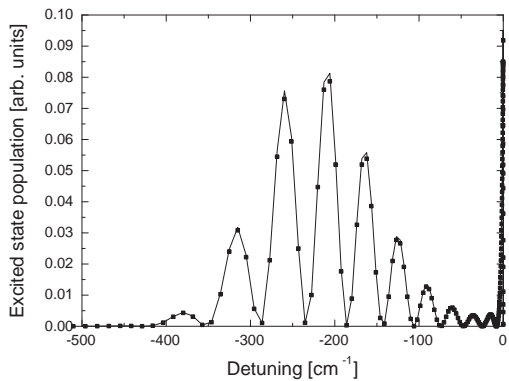}
\caption{Comparison between the predicted excited state level populations
  after the pump pulse from the model (solid line) and the results of the
  quantum dynamical calculation (squares).}
\label{AnaEndPop}
\end{figure}

In order to check the validity of \eref{PAtargetrate} in the perturbative
regime, a model study is performed. The excitation of a vibrational wavepacket
in the excited state, $\psi_e(R;t)$ is calculated with \eref{pumpsuperp} for a
single Fourier-limited Gaussian pump pulse. The detuning of the pulse is set
to $\Delta = 251$\,cm$^{-1}$ and the temporal intensity width to 100\,fs
FWHM. These values match the experimental conditions of Ref.\
\cite{WenzelSeinPaper}. In \fref{AnaEndPop}, the projections of the excited
state wavepacket onto the excited vibrational eigenfunctions $\varphi^e_v(R)$
are shown as a function of the detuning, $E^e_v$ (solid line). This is
compared in \fref{AnaEndPop} to the non-perturbative calculation for the same
pulse parameters (squares). Good agreement of the projections is observed in
the range of resonant levels.

\begin{figure}[tbp]
\center
\includegraphics[width=0.6\textwidth]{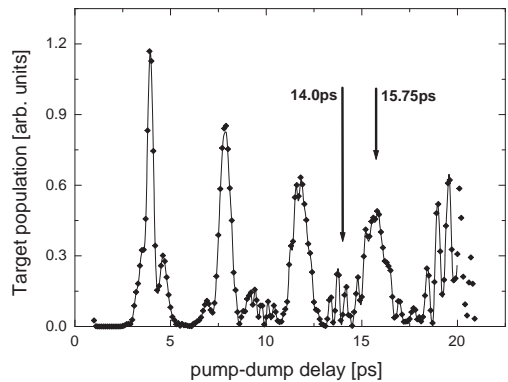}
\caption{Target state population $|b_w|^2$ for $w=119$ from
  \eref{PAtargetrate} (solid line) compared to the results of the quantum
  dynamical calculation (squares). The peaks are closely connected to the
  wavepacket dynamics in the excited state potential well. The temporal
  smearing out of peak heights is due to anharmonic dispersion of the
  wavepacket.  }
\label{PumpdumpUnshaped}
\end{figure}

\begin{figure}[tbp]
\center
\includegraphics[width=0.7\textwidth]{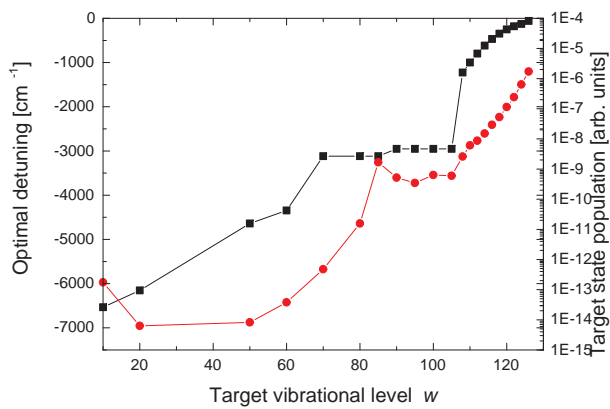}
\caption{Optimal detuning (black \fullsquare, left axis) for pump-dump
  photoassociation with two identical unshaped femtosecond laser pulses,
  estimated from \eref{PAtargetrate} for a range of target vibrational levels
  $w$ in the ground state. The maximum attainable population of these levels
  (red \fullcircle) is shown on the right axis.  }
\label{generaldata}
\end{figure}

As described above, the perturbative pump-dump model is only valid for
non-overlapping pulses, because transient dynamics occurring during the pulsed
excitation are not included. Therefore, in a second step it is checked if
\eref{PAtargetrate} holds for two-pulse pump-dump processes. Assuming two
pulses with width and detuning, as given above, the pump-dump process is
calculated with \eref{pumpsuperp3} for varying delay time. This yields the
ground state wavepacket which is then projected onto a single target
vibrational level. The level $w=119$, bound by $E^g_{119}=$6.86\,cm$^{-1}$, is
selected, because it features a large Franck-Condon factor from the excited
state vibrational levels that are populated at the given detuning of the laser
pulse (see \sref{subsec:optdet} and \fref{generaldata}, below). The population
in this target level is shown in \fref{PumpdumpUnshaped} (solid line) as a
function of the pump-dump delay time. It is compared to the results obtained
by solving the time-dependent Schr{\"o}dinger equation (squares). The
agreement between the perturbative model and the full quantum dynamics with a
non-perturbative treatment of the field is convincing.

Note that in the present case, the optimal pump-dump delay is given by the
full excited state wavepacket round-trip time $T_\mathrm{vib}$ and integer
multiples thereof. According to the Franck-Condon principle, this indicates
that the dump transition occurs at the outer turning point of the excited
state wavepacket, as indicated in \fref{PumpDumpIntroduction}. This is not
necessarily the case, since it depends strongly on the molecular potential
curves and the laser detuning. The dump transition will occur at half-integer
multiples of $T_\mathrm{vib}$, corresponding to the inner turning point if the
potential facilitates a deceleration mechanism at short distances. This is the
case for a soft repulsive wall such as in the outer well of the
0$_g^-(p_{3/2})$ state \cite{PumpDumpKoch}, or for resonant coupling
\cite{ChristianeD1}.

\subsection{Optimal detuning}
\label{subsec:optdet}

In order to find a window of detunings suitable to populate target levels in
the ground state, \eref{PAtargetrate} is evaluated for unshaped
Fourier-limited Gaussian pulses of 100\,fs FWHM intensity width. For a range
of delay times encompassing several round-trip times, the Franck-Condon matrix
$M^{vv^\prime}_w(\Delta t)$ is calculated for each ground state target level
$w$.  The optimal detuning is inferred from the maximum of the main diagonal
of $M^{vv^\prime}_w(\Delta t)$ and the corresponding maximum target state
population is calculated.

\Fref{generaldata} shows the result for both the optimal detuning and
corresponding target state population as a function of the vibrational level
$w$ of the ground electronic state. A strong dependence of the maximum
attainable population on the vibrational level is observed spanning eight
orders of magnitude. The largest target populations are found for small laser
detunings and correspond to high target vibrational levels. This reflects the
large free-bound Franck-Condon factors for highly excited vibrational
levels. It also shows that the bound-bound Franck-Condon factors for the dump
step favor high target levels for small detunings.

\subsection{Adaptive feedback control}
\label{subsec:control}

With the perturbative model, optimization calculations are carried out for
pump-dump photoassociation. As in the previous section, the pump and dump
pulses are assumed to be Fourier-limited Gaussian pulses with 100\,fs FWHM
intensity width and $\Delta = 251$\,cm$^{-1}$ detuning. The goal for the
optimization is to find the pump pulse that maximizes the population in the
$w=119$ ground state vibrational level, which is a suitable level for the
chosen detuning due to its large Franck-Condon overlap (see
\sref{subsec:valid}). The spectral components of the pump pulse are optimized
with respect to their phase and amplitude. The pump-dump delay time, which
represents a parameter in these calculations, is set either to 15.75\,ps or to
14.0\,ps: As shown in \fref{PumpdumpUnshaped}, for $\Delta t=15.75\,$ps a
maximum is observed in the target level population of the 'natural' dynamics,
i.e. the dynamics obtained with an unshaped pump pulse. On the other hand,
$\Delta t=14.0\,$ps represents a delay for which the natural dynamics leads to
almost no target level population.

\begin{figure}[tbp]
\center
\includegraphics[width=0.7\textwidth]{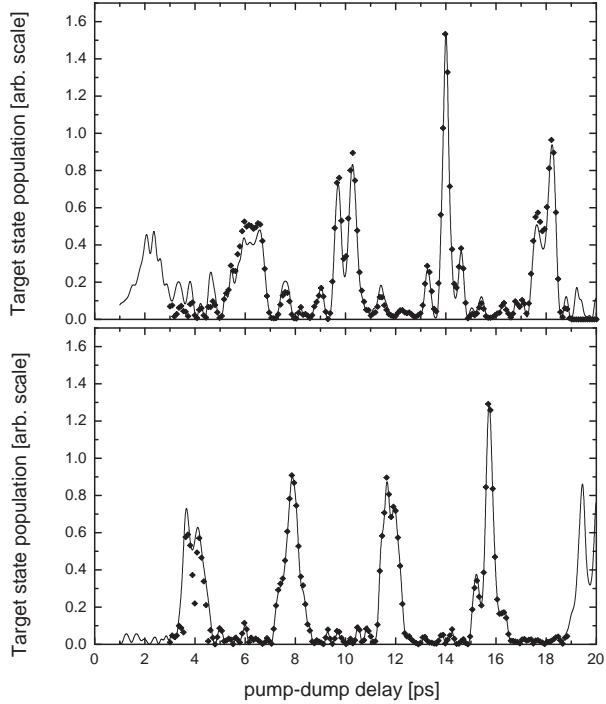}
\caption{Relative target state ($w=119$) population versus pump-dump delay for
  optimized pump pulses. The target level population on the vertical axis of
  the graphs is given in the same units as in \fref{PumpdumpUnshaped} and can
  thus directly be compered. The optimizations are carried out with a delay
  parameter of 14.00\,ps (upper trace) and 15.75\,ps (lower trace),
  corresponding to a minimum and a maximum in the 'natural' dynamics after an
  unshaped pump pulse, cf. \fref{PumpdumpUnshaped}. The pulse parameters are
  the same as in \fref{AnaEndPop}. The prediction of the model
  \eref{PAtargetrate} (solid line) is compared to the solution of the
  time-dependent Schr\"odinger equation (squares). Note the deviations
  occurring at small delay times which are due to transient excitations.  }
\label{PumpdumpShaped}
\end{figure}

The resulting target state population with optimized phases and amplitudes is
shown as a function of the pump-dump delay time in \fref{PumpdumpShaped}
(solid lines). Note that the target level population on the vertical axis of
both graphs in \fref{PumpdumpShaped} and of the graph in
\fref{PumpdumpUnshaped} is given in the same units and can thus directly be
compered. In \fref{PumpdumpShaped}, it is clearly seen that for both
optimization calculations the maximum population is indeed obtained at the
delay time for which the pump pulse is optimized. For $\Delta t=14.00\,$ps,
the dynamics is modulated such that an efficient dump transition may occur at
delay times when almost no target level population is obtained with an
unshaped pump pulse, see the left arrow in \fref{PumpdumpUnshaped}. In this
case, the target level population for the optimized pulse amounts to 1.6
arb. units., whereas less than 0.1 arb. units are observed for the same
pump-dump delay with an unshaped pulse, corresponding to an increase by
increased by more than a factor of 10. This illustrates that the pulse shaper
can be used to a great extent to control the wavepacket dynamics. The optimal
pulse shapes also lead to a higher maximum of the target level population as
compared to the maximum population achieved for the optimal pump-dump delay
time with unshaped pulses: for the optimization at $\Delta t=14.0\,$ps, the
target population is enhanced by about 40\%. For 15.75\,ps a smaller
enhancement of about 10\% is observed. The two different levels of
optimization are attributed to the different integrated energies of the
obtained optimal pulse, as detailed below.

The results of the optimization are compared with the results of the
time-dependent Schr{\"o}dinger equation using the optimal pulses, obtained
above, as pump pulses. The time delay between the pump and dump pulses is
varied to investigate whether the optimal result is indeed obtained for the
intended delay and how in general the pulse shapes affect the excited state
wavepacket dynamics. As shown in \fref{PumpdumpShaped}, these results agree
well with the model, except for very short delay times where the pulses still
overlap.

\begin{figure}[tbp]
\center
\includegraphics[width=0.9\textwidth]{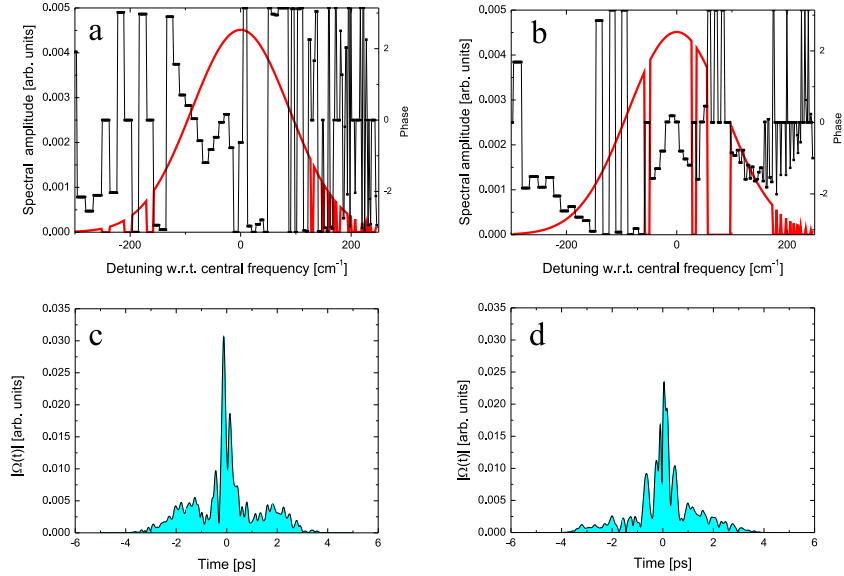}
\caption{Upper panels: Amplitude (red \full) and phase (black \dotted) mask
  patterns of the pulse shaper to generate the optimized pulses for $\Delta t
  =14.00\,ps$ (a) and $\Delta t=15.75\,$ps (b). The amplitudes are multiplied
  by the spectral envelope of the pulse. Note the chirp-like structures in the
  phases of the two pulses. Lower panels: Temporal amplitude of the optimized
  pulses for the 14.00ps case (c) and the 15.75ps case (d).  }
\label{OptimalPulses}
\end{figure}

The mask patterns of the pulse shaper for amplitude (\full) and phase
(\dotted) are shown for the two preset delay times in the top panels (a) and
(b) of \fref{OptimalPulses}. The spectrum contains frequencies for which the
excited state levels $v=204$ to $v=249$ are resonantly excited from the
initial state.  For $\Delta t=14.0\,$ps, \fref{OptimalPulses} (a), almost the
full spectrum is retained, whereas for $\Delta t=15.75\,$ps,
\fref{OptimalPulses} (b), some frequency components are removed from the pulse
spectrum. This leads to a smaller integrated energy of the optimized pulses
for the $\Delta t=15.75\,$ps optimum. Therefore a smaller target level
population is observed on maximum (about 1.3 arb. units) for the 15.75\,ps
calculation as compared to the 14.0\,ps calculation (1.6
arb. units). Chirp-like structures in the spectral phase emerge in both
cases. In other runs of the optimization algorithm, such structures have been
obtained as well.

\begin{figure}[tbp]
\center
\includegraphics[width=0.8\textwidth]{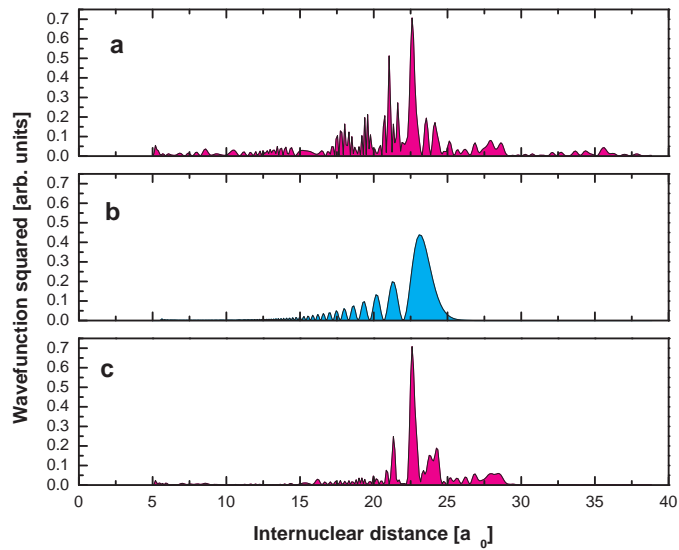}
\caption{Excited state wavepackets in position space at
  $t=15.75\,$ps after the pump pulse maximum.
  The upper panel (a) shows the
  wavepacket obtained after an unshaped pulse, while
  the lower (c) panel displays the
  wavepacket resulting from excitation with a pulse optimized
  for $\Delta t=15.75\,$ps. To estimate the dump efficiency,
  these wavepackets have to be compared to the stationary
  ground state wavefunction of the target level
  shown in the middle panel (b). Note the reduced
  probability density at small internuclear distances of the
  wavepacket after the optimized pulse as
  compared to the unshaped case.
}
\label{Wavefunctions}
\end{figure}

The temporal profiles of the optimized pulses are shown in the bottom panels
(c) and (d) of \fref{OptimalPulses}. The pulses are stretched in time compared
to the time duration of the unshaped pulses of 100\,fs FWHM. For $\Delta
t=14.0\,$ps, \fref{OptimalPulses} (c), three sub-pulses are observed which are
separated by about 2\,ps. This has to be compared to the vibrational times of
the excited state levels which range from 2.7\,ps for $v=204$ to 10.0\,ps
for $v=249$ and 4\,ps for the central level $v=219$ resonant with the carrier
frequency. In the Wigner representation of the optimal field (data not shown),
an interference pattern between the three sub-pulses is observed indicating
the intricate phase relation between the sub-pulses. The 'holes' in the
spectrum observed for $\Delta t=15.75\,$ps, as shown in \fref{OptimalPulses}
(b), correspond mostly to transition frequencies to levels with long
vibrational periods (for example, $v=227$ to $v=233$ with $T_\mathrm{vib} =
5.1\,$ps to 6.3\,ps).

Certain wavepacket components are filtered out to facilitate a shaping of the
excited state wavepacket which is favorable for the dump step. This is
illustrated for $\Delta t=15.75\,$ps in \fref{Wavefunctions}. The excited
state wavepacket at $t=15.75\,$ps after the unshaped, \fref{Wavefunctions}
(a), and optimized, \fref{Wavefunctions} (c), pump pulse is compared to the
vibrational wavefunction of the ground state target level,
\fref{Wavefunctions} (b). Optimal dump conditions are obtained when the
wavepacket is focused in the region where the target state wavefunction
possesses its outermost maximum. The optimization leads to a removal of parts
of the wavepacket in the inner region which would cause destructive
interference in the dump process.

\section{Conclusions and outlook}
\label{sec:concl}

We have investigated to which extent the formation of ultracold molecules by
pulsed photoassociation can be controlled using adaptive feedback control and
shaped femtosecond laser pulses. For this purpose we have set up a
perturbative pulse shaper model and combined it with a genetic algorithm. This
concept allows for a numerically efficient optimization with realistic
conditions for the femtosecond laser pulse and the pulse shaper. Using the
model for a pump-dump scheme, the population in a specific target level in the
ground electronic state is optimized. A considerable enhancement factor of
about 40\% is achieved compared to highest population achieved with unshaped
pulses, while keeping the pulse duration as short as possible. The maximum
pump-dump enhancement in the perturbative, low-energy regime, is determined by
the Franck-Condon factors of the two electronic transitions. Significantly
larger enhancements may only be expected for high-intensity pulses beyond the
perturbative regime, where Rabi cycling and dynamic Stark shifts occur.

The adaptive feedback optimization has shown that for different time delays
between the pump and dump pulses, a maximum in the ground state molecule
formation rate of similar height is achieved. This implies that the pump-dump
delay represents an uncritical parameter as long as it is shorter than the
spontaneous decay time, in the sense that the optimization algorithm may find
an optimal solution with a comparable target level population for arbitrary
delay times. The analysis of the excited state wavepacket dynamics reveals
that the shaped pulse induces a spatial focusing of the wavepacket in the
Franck-Condon region of the dump pulse at the specified pump-dump delay. The
pulse shaper can therefore greatly alter the dynamics of the wavepacket to
optimize ground state molecule formation.

The model described in this work is advantageous compared to standard
theoretical approaches such as optimal control theory
\cite{RabitzRapidlyConvergent} in that it closely resembles the situation in
experiments. Optimal control calculations generally converge faster than
genetic algorithm optimizations and extremely efficient pulses can be obtained
\cite{VibrationalCoolingKoch}. However, these pulses are usually too complex
to be directly implemented in an experiment
\cite{HornungPRA02a,MancalCPL02}. In this work, the optimized pulses are
obtained in the frequency domain and are given in terms of the transmission
coefficients and phase shifts of a pulse shaper. The optimization results can
therefore directly by be employed in experiments.

Future improvements of the presented calculations may follow several lines. To
obtain absolute rates for molecule formation, a thermal average needs to be
performed. This will be important to find pulse shapes for which association
dominates over photodissociation of ultracold molecules
\cite{WenzelSeinPaper,brown2006:prl}. For the collision energies in a
magneto-optical trap, such an average requires inclusion of partial waves with
$l>0$. Second, all relevant electronic states need to be incorporated in the
model. In particular, resonant spin-orbit coupling is known to lead to
perturbations in the excited state vibrational spectrum
\cite{AmiotPRL99,JelassiPRA06} which might be important for ground state
molecule formation \cite{ChristianeD1}. Furthermore, it will be interesting to
investigate the influence of the laser field polarization on ground state
molecule formation, in particular to find shaped pulses which excite specific
molecular symmetries. Finally, higher order terms may be included in the
perturbative derivation of the pulse shaper model to allow for investigation
of the nonlinear excitation regime. Here, higher enhancement factors beyond
the Franck-Condon regime may be expected. Based on the findings in this work,
dedicated experiments on pulsed photoassociation of ultracold rubidium
molecules are in progress.

This work has been supported by the Deutsche Forschungsgemeinschaft in the
frame of the Schwerpunktprogramm 1116 and by the European Commission in the
frame of the Cold Molecule Research Training Network under contract
HPRN-CT-2002-00290.  U. P. acknowledges support by the ESF network
``Collisions in Atom Traps''. C.P.K. acknowledges financial support from the
Deutsche Forschungsgemeinschaft.

\section*{References}


\end{document}